\documentclass{PoS}
\usepackage{textcomp, amsmath, amssymb, bbm}
\newcommand{\V}{\boldsymbol{V}}

\title{A $U(2)^3$ flavour symmetry in Supersymmetry}

\ShortTitle{A $U(2)^3$ flavour symmetry in Supersymmetry}

\author{\speaker{Filippo Sala}\thanks{Based on \cite{Barbieri:2011ci, Barbieri:2011fc}.}\\
        Scuola Normale Superiore and INFN, Piazza dei Cavalieri 7, 56126 Pisa, Italy\\
        E-mail: \email{filippo.sala@sns.it}}

\abstract{A
$U(2)^3$ flavour symmetry acting on the first two generations of quarks
partially explains the
hierarchies of the yukawa couplings, and provides a natural embedding for Supersymmetry with heavier
first two generations, where collider constraints are not in conflict with the requirement of naturalness
and the SUSY CP problem is solved. Within this context a specific pattern of flavour symmetry breaking
is considered.
The $K$, $B_d$ and $B_s$ mixing amplitudes show a definite correlation that can resolve existing tensions in the
CKM fit, pointing in this way to sbottom and gluino masses below about 1.5 TeV.
Potentially sizeable contributions to both indirect and direct CP violation in $B$ decays are allowed, even
in the absence of flavour-blind phases.
In case some effects are observed, the peculiar pattern in $\Delta F = 2$ and $\Delta B = 1$ observables
may allow to distinguish between this and other models.
}

\FullConference{Proceedings of the Corfu Summer Institute 2011 School and Workshops on Elementary Particle Physics and Gravity\\
		September 4-18, 2011\\
		Corfu, Greece}

\begin{document}

\section{Introduction and motivations}

The Standard Model (SM) description of flavour and CP violation, encoded in the Cabibbo Kobayashi Maskawa (CKM) matrix,
has shown an excellent agreement with data so far.
If one wants to allow for New Physics (NP) near the Fermi scale, as required by naturalness, one then has to explain
why no large deviations from the above flavour picture show up in experiments.\\
A possible approach to solve this issue is assuming that beyond the SM physics respects some flavour symmetry.
The most popular attempt in this direction is the Minimal Flavour Violation (MFV) paradigm
{\cite{Chivukula:1987py, Hall:1990ac, D'Ambrosio:2002ex}}: NP is formally invariant under a
$U(3)^3 = U(3)_{q_L} \times U(3)_{u_R} \times U(3)_{d_R}$ symmetry,
broken only by the SM Yukawas, which are promoted to spurion fields transforming as $Y_u \sim (3,\bar{3},1)$ and
$Y_d \sim (3,1,\bar{3})$.
While this makes possible, on general model independent grounds, to lower the scale of NP to a few TeV
\cite{Isidori:2010kg}, it lacks of an explanation of
(i) the smallness of the measured electric dipole moments, which would be generated by flavour-blind CP phases,
(ii) the hierarchy of the fermion masses and the specific pattern of mixing angles.\\
The first issue can be accomodated e.g. in Supersymmetry with heavier first two generation sfermions \cite{Cohen:1996vb,
Barbieri:2010pd, Barbieri:2010ar, Barbieri:2011vn}.
This scenario is favoured by current LHC sparticle searches if one wants to preserve the Supersymmetric solution to the hierarchy problem \cite{Papucci:2011wy}.
On the other hand, an attempt in the direction of explaining the hierarchies of the fermions is reducing the $U(3)^3$
symmetry to a $U(2)$ acting on the first two generations of quarks, irrespective of their chirality
\cite{Pomarol:1995xc, Barbieri:1995uv}.
However it turns out that this symmetry is not enough to suppress right-handed currents contribution to the $\epsilon_K$ parameter
\cite{Barbieri:1997tu, Barbieri:2011ci}.
The above considerations motivate us to study the consequences of a $U(2)^3 = U(2)_{q_L} \times U(2)_{u_R} \times U(2)_{d_R}$
flavour symmetry in the context of Supersymmetry with hierarchical squark masses.

\section{The $U(2)^3$ framework within Supersymmetry}
In the limit of exact $U(2)^3$ the up, down, strange and charm quark masses are zero and the CKM matrix $V_{CKM}$ is the identity.
Our choice for the breaking of this symmetry is dictated by minimality: we introduce two spurions $\Delta Y_u \sim (2,\bar{2},1)$ and $\Delta Y_d \sim (2,1,\bar{2})$ to generate the masses and mixings of the first two generation quarks, and a spurion $\V \sim (2,1,1)$ to let the first
two generations communicate with the third one.
By sole $U(2)^3$ transformations the spurions can be set to the form
\begin{equation}
\V^T = (0, \epsilon), \quad \Delta Y_u = R^u_{12} \cdot \Delta Y_u^{diag},
\quad \Delta Y_d = \Phi \cdot R^d_{12} \cdot \Delta Y_d^{diag},
\end{equation}
where $\epsilon$ is a parameter of the order of $|V_{cb}|$, $R^{u,d}_{12}$ are rotations in the $1 2$ sectors,
$\Phi = $ diag$(e^{- i \varphi},1)$, with $\varphi$ giving rise to the CKM matrix phase.
At first order in the spurions the Yukawa matrices read
\begin{equation}
Y_u = y_t \left( \begin{array}{c|c}
\Delta Y_u \; & x_t \V\\
\hline
 0            & 1
\end{array}
\right), \qquad
Y_d = y_b \left( \begin{array}{c|c}
\Delta Y_d \; & x_b \V \\
\hline
 0            & 1
\end{array}
\right),
\end{equation}
so that they are formally invariant under $U(2)^3$.
Analogously, we assume the flavour symmetry breaking in the soft squark masses and $A$ terms to be controlled by the same spurions\footnote{This spurion structure is effectively realized in MFV at large $\tan \beta$ \cite{Feldmann:2008ja,Kagan:2009bn}. A $U(2)^3$ symmetry, broken only by the $A$ terms and without this specific spurion structure, has also been considered in the context of radiative flavour violation in the MSSM in \cite{Crivellin:2008mq,Crivellin:2011sj}.}.
After rotating to the mass basis for both the quarks and their superpartners, $V_{CKM}$ and the supersymmetric mixing matrices appearing in the vertices
$\left( \bar{d}_{L,R} \, W^{L,R} \, \tilde{d}_{L,R} \right) \tilde{g}$ take the correlated forms
\begin{equation}
 V_{CKM}=\left(\begin{array}{ccc}
 1- \lambda^2/2 &  \lambda & s_u s e^{-i \delta}  \\
-\lambda & 1- \lambda^2/2   & c_u s  \\
-s_d s \,e^{i (\varphi+\delta)} & -s c_d & 1 \\
\end{array}\right), \quad
W^L = \left(\begin{array}{ccc}
 c_d &  \kappa^* & - \kappa^* s_L e^{i \gamma}  \\
- \kappa  &  c_d & -c_d s_L e^{i \gamma}   \\
  0  & s_L e^{-i \gamma} & 1
\end{array}\right), \quad
W^R = \mathbbm{1},
\end{equation}
where $\kappa = s_d e^{i (\delta + \varphi)}$, the phases $\delta$ and $\varphi$ are related to each other and to the other parameters
via $s_u c_d - c_u s_d e^{- i \varphi} = \lambda e^{i \delta}$, the new parameter $s_L > 0$ is of order $\lambda^2 \sim \epsilon$,
like $s$, and $\gamma$ is an independent CP violating phase, which is zero if we do not allow for phases outside the spurions.
$W_R$ is equal to the identity to an accuracy which is sufficient to avoid a too large contribution to $\epsilon_K$\footnote{
Note that a non-minimal breaking pattern (e.g. with $V_u \sim (1,\bar{2},1)$ and $V_d \sim (1,1,\bar{2})$ \textit{instead of} $V$)
would in general spoil this feature. For a recent discussion of the inclusion of these two spurions \textit{in addition to} $V$ see \cite{Barbieri:2012bh}.}.

\section{Phenomenology of $\Delta F = 2$ and $\Delta B = 1$ observables}
The mixing amplitudes in the kaon and $B_{d,s}$ systems,
including SM and gluino-mediated contributions,
read:
\begin{equation}
M_{12}^K  = (M_{12}^K)^\text{SM, tt} \left(1 + |\xi_L|^4 F_0\right) + (M_{12}^K)^\text{SM, tc + cc},\qquad
M_{12}^{B_{d,s}} = (M_{12}^{B_{d,s}})^\text{SM} \left(1+\xi_L^2 F_0\right),
\end{equation}
where $\xi_L = (c_d s_L/|V_{ts}|)\, e^{i \gamma}$ and $F_0(m_{\tilde{b}}, m_{\tilde{g}})$ is a positive function of
the sbottom and gluino masses. In this way one obtains correlated
expressions for the indirect CP violating parameter $\epsilon_K$ and the mixing-induced CP asymmetries in
$B^0 \to \psi K_S$ and $B^0 \to \psi \phi$ decays:
\begin{equation}
 \begin{array}{rl}
\epsilon_K &= \epsilon^\text{SM, tt}_K \left(1 + |\xi_L|^4 F_0\right) +
\epsilon^\text{SM, tc + cc}_K, \vspace{.2 cm}\\
S_{\psi K_S} &=\sin\left(2\beta + \phi_\Delta\right), \vspace{.2 cm}\\
S_{\psi\phi} &=\sin\left(2|\beta_s| - \phi_\Delta\right),
\end{array}
\end{equation}
where $\beta$ and $\beta_s$ are the SM mixing phases and $\phi_\Delta=\text{arg}\!\left(1 + \xi_L^2 F_0\right)$.\\
The above effects are interesting in light of the tension in the CKM description of flavour and CP violation,
namely among $\epsilon_K$, $S_{\psi K_S}$ and $\Delta M_d/\Delta M_s$ \cite{Lunghi:2008aa, Buras:2008nn, Altmannshofer:2009ne,
Lunghi:2010gv, Bevan:2010gi}. This tension can be solved in our framework by the new contributions we obtain
to the first two observables.
Note that a specific prediction of the model (with the assumption of gluino dominance) is a definite sign of
the correction to $\epsilon_K$, which is the one needed in order to solve the CKM tension.\\
We then perform a global fit varying the SUSY parameters $\xi_L$, $F_0$ and $\gamma$ together with
the four parameters of the CKM matrix, using all the observables listed in Table 1 of \cite{Barbieri:2011ci} as constraints.
This leads, at the $90 \%$ C.L., to the preferred intervals
\begin{equation}
|\xi_L| \in [0.8, \, 2.1], \quad
\phi_\Delta \in [-9^\circ,\, -1^\circ], \quad
\gamma \in [-86^\circ, \,-25^\circ] \; \text{ or } \;\gamma \in [94^\circ, \,155^\circ], 
\label{DeltaF2}
\end{equation}
and to the predictions $S_{\psi \phi} \in [0.05, \,0.20]$ and  $m_{\tilde{g}}, \; m_{\tilde{b}} \lesssim 1.5$ TeV
(the latter due to $F_0 \neq 0$).
The favoured value for $S_{\psi \phi}$ is within the $1 \sigma$ interval of the recent LHCb result \cite{LHCb:2012moriond}.\\
It should be noted that some of the predictions we make are independent of the specific model under consideration
(Supersymmetry with hierarchical squark masses),
namely (i) the absence of a new phase in $M_{12}^K$,
(ii) the presence of a new phase in $B_{s,d}$ mixing
and (iii) the universality $M_{12}^{B_d} = M_{12}^{B_s}$.

\begin{figure}[tbp]
\begin{center}
\includegraphics[width=7 cm]{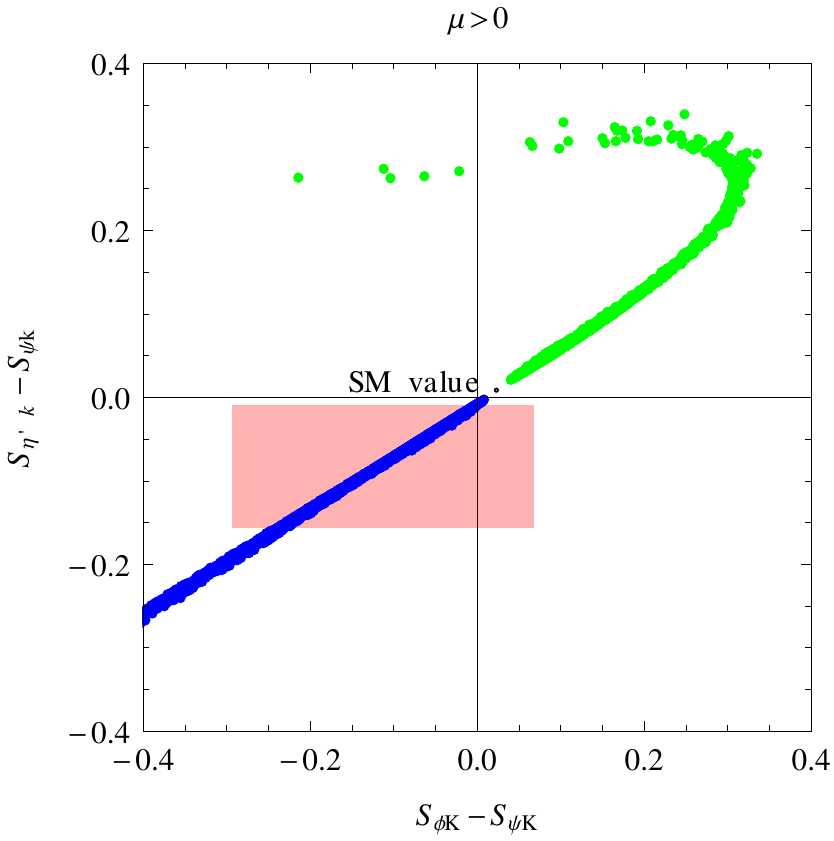}
\qquad
\includegraphics[width=7 cm]{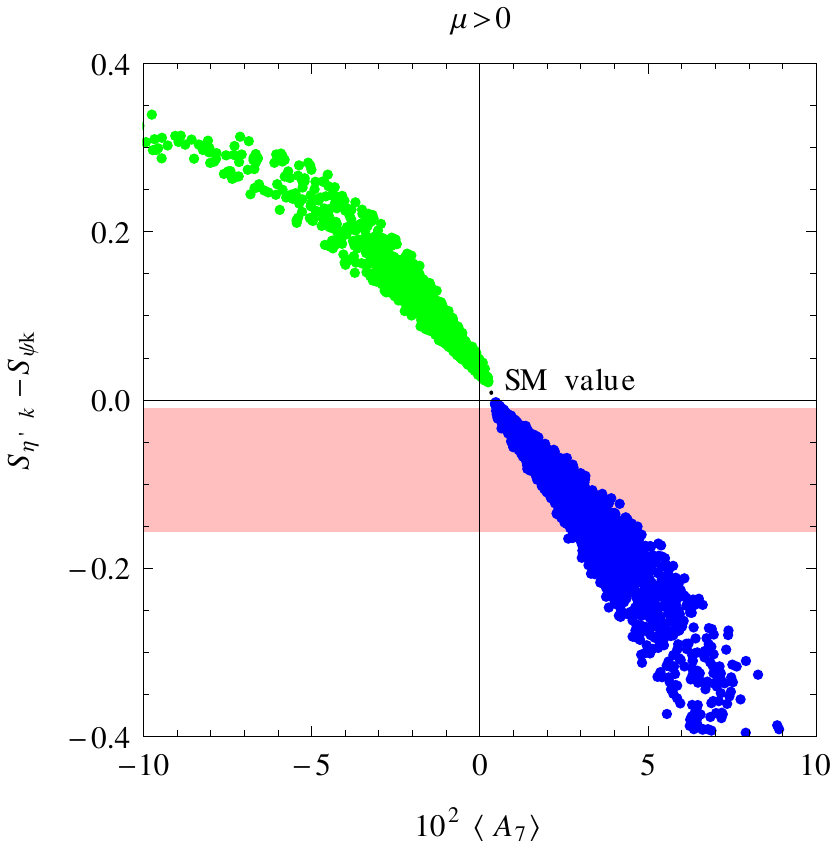}
\end{center}
\caption{Left: correlation between $S_{\phi K_S} - S_{\psi K_S}$ and $S_{\eta' K_S} - S_{\psi K_S}$.
Right: correlation between $\langle A_7 \rangle$ and $S_{\eta' K} - S_{\psi K}$. Points with $\gamma > 0$ ($\gamma < 0$)
are shown in blue (green). The red shaded regions show the $1 \sigma$ experimental range. The plots where the supersymmetric
parameter $\mu$ is chosen to be negative do not show qualitative differencies,
apart from the inversion of the $\gamma > 0$ and $\gamma < 0$ points}
\label{fig:correlations}
\end{figure}

The presence of the phase $\gamma \,$ leads to sizeable contributions to CP asymmetries in B decays.
So, in \cite{Barbieri:2011fc}, we study them in the case of negligible flavour-blind phases, in order
to isolate the genuine $U(2)^3$ effect and make the link with the $\Delta F = 2$ observables evident
(for this reason we also assume gluino-mediated amplitudes to dominate).\\
More specifically, we study the angular CP asymmetries in $B\to K^* \mu^+ \mu^-$, $\langle A_7 \rangle$ and
$\langle A_8 \rangle$ \cite{Altmannshofer:2008dz}, and the CP asymmetries in $B \to \eta' K_S$ and $B \to \phi K_S$ decays,
$S_{\eta' K_S}$ and $S_{\phi K_S}$. The former turn out to be strictly correlated, $\langle A_8 \rangle \simeq - 
0.56 \langle A_7 \rangle$, confirming the result of \cite{Barbieri:2011vn}.
The latter take the form $S_f = \sin(2 \beta + \Phi_{\Delta} + \delta_f)$, where $\delta_f$
is a parameter encoding the direct CP violation contributions. In order to effectively ignore the
modified $B$ mixing phase, we show in figure \ref{fig:correlations} their correlation after subtraction of $S_{\psi K_S}$.
The plots shown are obtained scanning the gluino mass between $0.5$ and $1$ TeV, the sbottom mass, the $\mu$ and the
$A$ term between 0.2 and 0.5 TeV and $\tan \beta$ between 2 and 10. The new flavour parameters $|\xi_L|$ and $\gamma$
are required to lie in the region where the CKM fit tension is reduced (i.e. we impose \eqref{DeltaF2}); the most effective
constraint from other observables comes from BR($B \to X_s \gamma$).
We also study the direct CP asymmetry $A_{CP}(B \to X_s \gamma)$, the contribution we find is sizeable
and shows a specific correlations with the other direct CP violating observables.
However the sensitivity of this observable to NP is spoiled by non-perturbative SM effects, which can saturate
the future expected experimental sensitivity \cite{Benzke:2010tq}.\\
It is interesting to note that the above correlations between the $\Delta B = 1$ observables are very similar to
those obtained in MFV \cite{Altmannshofer:2008hc, Altmannshofer:2009ne} or in effective MFV \cite{Barbieri:2011vn}
with flavour-blind phases, while in our setup flavour-blind phases were assumed to be zero. Another important difference is that in our case we expect new CP violating effects in $K$ and $B$ mixing, which are instead absent in (effective) MFV.

\section{Summary and outlook}
A suitably broken $U(2)^3$ flavour symmetry is considered as an alternative to MFV.
A model independent consequence is a universal shift in $B_s$ and $B_d$ mixings, which can be complex,
and a non standard real contribution to $\epsilon_K$.
The phenomenological consequences of this symmetry are explored in the context of Supersymmetry with
hierarchical squark masses.
The contributions to $\Delta F = 2$ observables are required to ameliorate tensions in the CKM fit,
in this way one obtains for the sbottom and gluino masses the preferred region $m_{\tilde{b}}, \;m_{\tilde{g}}
\lesssim 1.5$ TeV.
Even if flavour-blind phases are negligible, one finds potentially sizeable contributions to both indirect and direct
CP violation in $B$ decays (the processes studied are $B \to \psi \phi$, $B \to \eta' K_S$, $B \to \phi K_S$, $B \to K^* \mu^+ \mu^-$ and $B \to X_s \gamma$).
These effects, while being consistent with data so far, can be in the region to be probed by LHCb and the next
generation $B$ factories, and display a peculiar pattern in $\Delta F = 2$ and $\Delta B = 1$ observables that may
help in distinguishing other models form the one presented here, in case some signal is observed.\\
The interest in studying this flavour symmetry goes beyond the specific case of Supersymmetry, see \cite{Barbieri:2012uh}
for a recent treatment of $U(2)^3$, both in a model independent way and in composite Higgs models, and for a possible
extension to the lepton sector.

\acknowledgments
I thank the organizers of the Corfu Summer Institute 2011 \textit{School and Workshops on Elementary
Particle Physics and Gravity} for the very kind hospitality and for giving me the opportunity to give a talk.
I also thank P. Campli and D. Straub for comments on the manuscript and R. Barbieri for involving me in this project.

\end{document}